\documentclass[aps,prd,showpacs,preprintnumbers,twocolumn]{revtex4}
\usepackage{epsfig}
\usepackage{latexsym,amsmath,amssymb,amstext}
\begin{document}
\title{Thermodynamics of free Domain Wall fermions }
\author{R.\ V.\ \surname{Gavai}}
\email{gavai@tifr.res.in}
\affiliation{Department of Theoretical Physics, Tata Institute of Fundamental
         Research,\\ Homi Bhabha Road, Mumbai 400005, India.}
\author{Sayantan\ \surname{Sharma}}
\email{ssharma@theory.tifr.res.in}
\affiliation{Department of Theoretical Physics, Tata Institute of Fundamental
         Research,\\ Homi Bhabha Road, Mumbai 400005, India.}

\begin{abstract}
Studying various thermodynamic quantities for the free domain wall fermions
for both finite and infinite fifth dimensional extent $N_5$, we find that
the lattice corrections are minimum for $N_T\geq10$ for both energy density
and susceptibility, for its irrelevant parameter $M$ in the range
1.45-1.50.  The correction terms are, however, quite large for small
lattice sizes of $N_T\leq8$. We propose modifications of the domain wall
operator, as well as the overlap operator, to reduce the finite cut-off
effects to within 10\% of the continuum results of the thermodynamic
quantities for the currently used $N_T=$6-8 lattices. Incorporating
chemical potential, we show that $\mu^2$ divergences are absent for a large
class of such domain wall fermion actions although the chiral symmetry is
broken for $\mu\neq0$.
\end{abstract}
\preprint{TIFR/TH/08-51}
\maketitle

\section{ Introduction}

The nature of QCD matter at high temperatures and densities has been of
interest due to the experiments at the Relativistic Heavy Ion
Collider(RHIC) in BNL, New York and the upcoming Large Hadron Collider(LHC)
in CERN, Geneva. Theoretically, it is expected that the spontaneously
broken chiral symmetries at low temperatures are restored at high
temperatures. While lattice methods have been very effective in predicting
the transition temperature as well as the nature of the the strongly
interacting matter near the transition temperature, fermions with exact
chiral symmetry on lattice are important for such a study of the chiral
phase transition. Kaplan\cite{kaplan} proposed to define fermions with
exact chiral symmetry on a five-dimensional (5-D) lattice with a mass term
$M$ in the form a step function(domain wall) and with an infinite extent
along the fifth dimension. The massless 4-D fermions are obtained as
localized on the wall, and are hence known as the domain wall fermions. On
a finite lattice needed for numerical simulations, however, fermions of
both chiralities exist with an exponentially small overlap between the
respective chiral states\cite{vranas}.  Currently, the most popularly used
fermions in QCD simulations at finite temperatures/densities are the
staggered fermions which have only a remnant chiral symmetry on the
lattice.  Moreover, they explicitly break spin and flavor symmetries.  The
full chiral symmetry for these fermions is recovered only in the continuum
limit, i.e., in the limit of vanishing lattice spacing.  In spite of the
(exponentially small in $N_5$, the number of sites in the fifth dimension)
chiral violation on the lattice, the domain wall fermions are more
promising than the staggered fermions due to their exact flavor and spin
symmetry on the lattice.  On the other hand these are more expensive to
simulate as the computational cost increases linearly with $N_5$. One has
to optimize $N_5$ and $M$ for full QCD simulations. In order to gain
insights on ways to minimize the lattice cut-off effects we study various
thermodynamic quantities of free domain wall fermions as a function of $M$
and $N_5$ with an aim to optimize the irrelevant lattice parameters for
faster convergence to their continuum values.  We find that by adjusting
the domain wall height $M$ in the range $1.45-1.55$ rather than the
frequently used choice of $M=1.0$, a faster convergence to the continuum
results for both finite and infinite values of $N_5$ is achieved.  However,
the cut-off effects are seen to be quite large on small lattices with
temporal extent  of 4-6 where most of the current QCD simulations are being
done.  We therefore examine modifications of the domain wall, as well as
the overlap kernel to minimize such corrections for small lattice sizes.

The plan of the paper is as follows: In section II, we compute the energy
density of free domain wall quarks on the lattice analytically and verify
that it yields the correct continuum limit.  In section III, the same
quantity is computed numerically and the various lattice parameters for
which the convergence to the continuum is fastest are estimated. In section
IV, we repeat the calculations of energy density in the presence of
chemical potential and susceptibility and confirm that this optimum M-range
does not shift significantly.  In section V, we propose a method of
reducing the lattice cut-off corrections to thermodynamic quantities  on
small lattice sizes, computed using both the chiral fermions, namely the
domain wall at infinite $N_5$ and the overlap fermions. This helps in
faster convergence to the continuum results even for $M=1.0$.

\section{Energy Density of Domain wall fermions}

The domain wall fermions\cite{kaplan} in the continuum are defined on a 5D
space-time with the mass  term in the fifth dimension in form of a domain
wall $\phi(M)=M\tanh(s)$.  This helps in localizing a fermion of definite
chirality on the domain wall.  The domain wall operator in the continuum is
given as,
\begin{equation}
\label{eqn:dwc}
D_{DW}=\sum_{\mu=1}^{4}\gamma_{\mu} \partial_{\mu}+\gamma_5 \partial_5+\phi(M).
 \end{equation}
The massless fermion modes in 4-D are obtained when the following
conditions are simultaneously satisfied.
\begin{eqnarray} 
\nonumber
 \sum_{\mu=1}^{4}\gamma_{\mu} \partial_{\mu}\psi=0\\ 
(\gamma_5 \partial_5+\phi(M))\psi&=&0.
\end{eqnarray}
It was shown that only one normalizable solution exist, bounded to the wall
at $s=0$ where the $\phi(M)$ changes abruptly. The corresponding analog of
the domain wall term on the lattice is of the form \begin{equation}
 \phi(M)=M\Theta(s)
\end{equation}
On the lattice we do not get a single massless mode by discretizing Eq.
(\ref{eqn:dwc}). This is because the lattice regulator is anomaly free, so
massless fermions of both handedness exist on the lattice.  A Wilson term
is needed to spatially separate the left and the right handed fermions in
the 5th dimension by localizing them on the domain wall and the anti-domain
wall respectively which are  separated from each other by the lattice
extent in the fifth dimension $N_5$.  To obtain thermodynamical quantities
of free fermions with exact chiral symmetry on the lattice in 4-D, we need
to divide out contribution of the heavy fermion modes which exist in the
fifth dimension. This is done by subtracting a pseudo-fermion
action\cite{shamir} from the standard 5-D action.  Following
Shamir\cite{shamir}, the domain wall fermion action on a $N^3 \times N_T
\times N_5 $ anisotropic lattice with lattice spacings of $a$, $a_4$ and
$a_5$ in the three spatial, the temporal and the fifth dimension
respectively can be written as, 
\begin{eqnarray}
\label{eqn:dwop}
\nonumber
&&S_{DW}=-\sum_{s,s'=1}^{N_5}\sum_{x,x'}\bar{\psi}(x,s)D_{DW}
(x,s;x',s',\hat{\mu},\hat{m})\psi(x',s')\\\nonumber
&&=-\sum_{s,s'=1}^{N_5}\sum_{x,x'}\bar{\psi}(x,s)\left[\left(\frac{a_5}{a} D_W(x,x')
+1\right)\delta_{s,s'}\right.\\
&&-\left.\left(P_{-}\delta_{s',s+1}+P_{+}\delta_{s',s-1}\right)\delta_{x,x'}
\right]\psi(x',s'),
\end{eqnarray}
with the boundary conditions
\begin{equation}
 P_{-}\psi_{N_5+1}=-\hat{m_q} P_{-}\psi_1,~~P_{+}\psi_{0}=-\hat{m_q} P_{+}\psi_{N_5}
\end{equation}
where $P_{\pm}=\frac{1\pm\gamma_5}{2}$ are the chiral projectors and
$\hat{m_q}$ is the bare quark mass in lattice units. $D_W$ is the
Wilson-Dirac operator defined on a 4-D lattice.  The volume of the system
is $V=N^3a^3$ and $T=1/(N_Ta_4)$ is its temperature. The chemical potential
$\mu a_4=\hat{\mu}$ is usually introduced as a Lagrange multiplier
corresponding to the conserved number density in the expression for the
Lagrangian. For the domain wall fermions we do not have a good prescription
for obtaining the conserved number density. Following Bloch and
Wetting \cite{wettig}, we incorporate the chemical potential in $D_W$ but
in a general form using the functions $K$ and $L$ \cite{gav} defined below.
These multiply the $1\pm \gamma_4$ factors in the Wilson-Dirac operator
leading to, 
\begin{eqnarray}
 \label{eqn:wilop}
\nonumber
&&D_W(x,x',\hat{\mu})=\left(3+\frac{a}{a_4}-M\right)\delta_{x,x'}-\\\nonumber
&&\sum_{j=1}^{3}\left(U_j^{\dagger}(x-\hat{j})\frac{1+\gamma_j}{2 }
\delta_{x,x'+\hat{j}}+U_j(x)\frac{1-\gamma_j}{2 }\delta_{x,x'-\hat{j}}\right)\\\nonumber
&&-\frac{a}{a_4}
\left(K(\hat{\mu})U_4^{\dagger}(x-\hat{4})\frac{1+\gamma_4}{2 }
\delta_{x,x'+\hat{4}}\right.\\
&&+\left.L(\hat{\mu})U_4(x)\frac{1-\gamma_4}{2 }\delta_{x,x'-\hat{4}}\right).
\end{eqnarray}
In this paper we consider the non-interacting fermions, i.e.,
$U_{\mu}(x)=1$.  Introducing $R$ and $\theta$ by
\begin{equation}
 \frac{K(\hat{\mu})+L(\hat{\mu})}{2}=R \cosh \theta ~~~
\frac{K(\hat{\mu})-L(\hat{\mu})}{2}=R \sinh \theta~,~
\end{equation}
the free Wilson-Dirac operator in Eq. (\ref{eqn:wilop}) can be
diagonalized in the momentum space in terms of the functions,
\begin{eqnarray}
\label{eqn:hdef}
h_j&=&\sin a p_j~,~h_4=-\frac{a}{a_4} R \sin(a_4p_4-i\theta)~,~\\\nonumber
h_5&=&M-\sum_{j=1}^{3}(1-\cos a p_j)
-\frac{a}{a_4}(1-R \cos (a_4 p_4-i\theta)).
\end{eqnarray}
such that
\begin{equation}
\label{eqn:wils}
D_W(\vec p,p_4) =-\sum_{i=1}^{4} i\gamma_i h_i-h_5.
\end{equation}
To study thermodynamics of fermions one has to necessarily take anti-periodic
boundary conditions along the temporal direction. Assuming periodic boundary
conditions along the spatial directions we obtain
\begin{eqnarray}
\nonumber
ap_j &=& \frac{2n_j\pi}{N},n_j=0,..,(N-1), j=1,2,3 ~{\rm and}\\
ap_4 &=&\omega_n= \frac{(2n+1)\pi}{N_T},n=0,..,(N_T-1)
\end{eqnarray}
It is to be noted that $M$, the height of the domain wall on the lattice, is
bound to $0<M<2$ to simulate one flavor quark on the lattice. To suppress
the heavy mode  contributions and recover a single chiral fermion,
pseudo-fermion fields are introduced which have the same action but with
$\hat{m_q}=1$ i.e with anti-periodic boundary condition in the fifth
dimension\cite{vranas}.  The fifth dimensional degrees of freedom can be
integrated out to yield an effective domain wall operator\cite{neu,eh} 
\begin{equation}
\label{eqn:dw}
 \frac{D_{DW}(\hat{m}_q)}{D_{DW}(1)}=1+\hat{m}_q+(1-\hat{m}_q)
\gamma_5\frac{1-T^{N_5}}{1+T^{N_5}},
\end{equation}
where the transfer matrix $T$ is 
\begin{equation}
\label{eqn:trmat}
 T=(1+\frac{a_5}{a}\gamma_5 D_W P_{+})^{-1}(1-\frac{a_5}{a} \gamma_5 D_W P_{-}).
\end{equation}
Since $T$ can be shown to be Hermitian for $\hat{\mu}=0$, and therefore has real
eigenvalues, $T^{N_5}$ has only positive eigenvalues for even $N_5$.
Introducing \cite{eh} a notation $|T|$, the function
$\frac{1-T^{N_5}}{1+T^{N_5}}$ in the domain wall operator can be expressed
in the form of a tanh function as in Eq. (\ref{eqn:tanh}).
\begin{equation}
\label{eqn:tanh}
 \frac{D_{DW}(\hat{m}_q)}{D_{DW}(1)}=1+\hat{m}_q-(1-\hat{m}_q)
\gamma_5\text{tanh}(\frac{N_5}{2}\ln |T|).
\end{equation}
  The above derivation of the effective domain wall
operator assumes that $1+T^{N_5}$ does not have any zero eigenvalues.
For if it does, then the contribution of the heavy modes is zero.  If
$\lambda$ be an eigenvalue of T, then this assumption requires that 
\begin{equation}
\label{eqn:cond}
\ln \lambda\neq i\frac{(2n+1)\pi}{N_5}~.
\end{equation}
This is clearly true for $\hat{\mu}=0$ for even the interacting fermions where $T$
is Hermitian and thus any $\lambda$ is real. However, once chemical
potential is introduced in the Wilson-Dirac operator, as above, $D_W$ and
$T$ are not Hermitian any longer for the free fermions themselves, leaving
open the possibility that this condition will not be met.

It is easy to see that three distinct limits are of interest in which we
should compute the various thermodynamic quantities for massless domain wall
operator. These are as follows: 
\begin{enumerate}
\item{$N_5\rightarrow\infty,a_5=\text{finite}$, where one obtains
exact chiral fermions for $\hat{m_q}=0$,}
\item{ $N_5\rightarrow\infty,a_5\rightarrow 0$, such that $L_5 = N_5 a_5$,
leading to an approximate form for the overlap fermions\cite{NeuNar}, and}
\item{$N_5=\text{finite},a_5=\text{finite}$, corresponding to the 
form of domain wall operator directly relevant for practical
simulations on the lattice}.
\end{enumerate}

\subsection{$N_5\rightarrow\infty,a_5=\text{finite}$}
In this limit, the tanh function in Eq. (\ref{eqn:tanh}) becomes sign
function and the resultant  effective domain wall operator is given as
\begin{equation}
\label{eqn:effdw}
D^{eff}_{DW}=1 +\hat{m}_q -(1 -\hat{m}_q ) \gamma_5\epsilon(\ln|T|)
\end{equation} 
For $\hat{m}_q =0$, this form of the domain wall operator satisfies the
Ginsparg-Wilson relation \cite{wil}, as shown in the appendix \ref{app1}.
Indeed, it is just like the overlap operator, but with a
different argument of the sign function. The finite size corrections to
various thermodynamic quantities computed with this lattice operator are
expected to be different from the overlap case.  For this type of Ginsparg
Wilson fermion too the introduction of chemical potential necessarily leads
to chiral symmetry breaking \cite{bgs} on the lattice because the action in
presence of $\hat{\mu}$ is not invariant under the chiral
transformations\cite{Luscher} on lattice.  Like in the case of the overlap
fermions, chiral symmetry is exactly realized for these wall fermions only
in the absence of chemical potential.

The energy density $\epsilon$ of the domain wall fermions in the chiral
limit is evaluated from the temperature partial derivative of the
partition function, $Z=\text{Det}(D^{eff}_{DW})$.  This is equivalent to
taking a partial derivative with respect to $a_4$ on a lattice of fixed
size $N_T$.  The energy density,
\begin{equation}
\label{eqn:edensity1}
 \epsilon = -\frac{1}{N^3 a^3 N_T }\left(\frac{\partial}{\partial a_4}\text{ln }Z
\right)_{a,\hat{\mu}}.
\end{equation}
can be evaluated analytically in terms of the quantities $q,s,t$ and
$s^{'},t^{'}$ defined below in Eq. (\ref{eqn:def1}), where the dash
denotes the $a_4$-derivative of the respective quantities. Defining
\begin{eqnarray}
 \label{eqn:def1}
\nonumber
h^2&=&\sum_{i=1}^{4}h_i^2~,~ s^2=h^2+h_5^2 ~,~\\\nonumber
t&=&s\sqrt{s^2-4h_5+4}~,~~q=s^2-2h_5+2~,~\\\nonumber
\alpha&=&\frac{\partial h_4}{\partial a_4}=\frac{a}{a_4^2} 
R\sin(a_4p_4-i\theta)~,~\\\nonumber
\gamma&=&\frac{\partial h_5}{\partial a_4}=\frac{a}{a_4^2}
(1-R\cos(a_4 p_4-i\theta))\\\nonumber
s^{'}&=&\frac{h_4\alpha+h_5\gamma}{s}~,\\
t^{'}&=&\frac{s^{'} t}{s}+\frac{s^2 (s s^{'}-2\gamma)}{t}
\end{eqnarray}
one has 

\begin{eqnarray}
\label{eqn:sgne}
\nonumber
&&\epsilon a^4=\frac{1}{N^3 N_T}\sum_{p_j,n}\left(\frac{2 t^{'}}{t}-
\frac{ 4 h_5 \gamma+4 s s^{'}(1+ s^2)}{2 h_5^2+2 s^2+s^4}\right.\\\nonumber
&&\left. \frac{+2 s s^{'} t+s^2 t^{'}-4 \gamma s^2-8 h_5s s^{'}
-2 \gamma t-2 h_5 t^{'}} {+s^2 t-4 h_5s^2-2 h_5 t}\right)\\
&&\equiv \frac{1}{N^3 N_T}\sum_{p_j,n}F(R,\omega_n,\vec{p}).
\end{eqnarray}
Setting $a=a_4$ after evaluating the $a_4$-derivatives, the summation
over the discrete Matsubara frequencies can be evaluated analytically by the
standard contour integral technique or numerically by explicitly summing
over them and the momenta $p_j$.   For the former, we need to determine
the singularities of the summand $F$ in Eq. (\ref{eqn:sgne}).  We outline
below briefly the results one obtains for the zero and finite temperature
cases.

\subparagraph{$T =0$, $\mu \ne 0$ :}
In order to obtain a general condition for eliminating the spurious
$\hat{\mu}^2$-divergences, we first calculate the energy density at zero
temperature in the limit $N_T\rightarrow\infty$ at finite a. The frequency
sum $1/N_T\sum_n$ in Eq. (\ref{eqn:sgne}) gets replaced by the integral
$\frac{1}{2\pi}\int_{-\pi}^{\pi}d\omega$ in this limit. Subtracting the
vacuum contribution corresponding to $\hat{\mu}=0$, i.e. $R = 1, \theta =0
$, the energy density at zero temperature is given by
\begin{equation}
 \epsilon a^4=\frac{1}{\pi N^3 }\sum_{p_j}\left[\int_{-\pi}^{\pi}F(R,\omega-i\theta)d\omega-\int_{-\pi}^{\pi}
F(\omega)d\omega\right]~.
\end{equation}
For brevity, we suppress from now on the explicit $p_j$-dependence of the
function $F$ although we retain the overall sign to remind of it.
\begin{figure}
\begin{center}
 \includegraphics[scale=0.6]{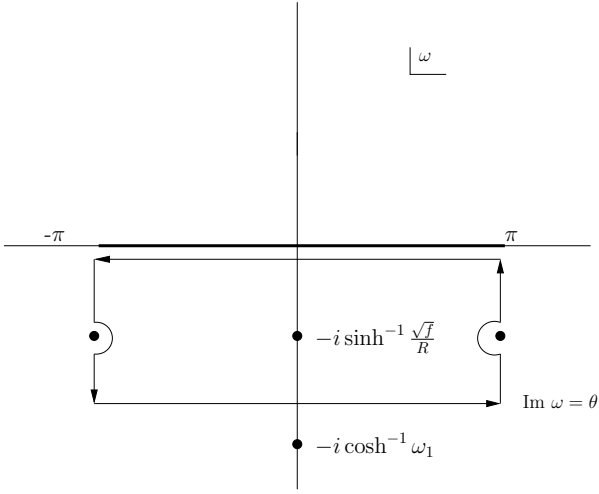}
\caption{Contour chosen for evaluating the energy density for nonzero value
 of chemical potential at zero temperature. The thick line
indicates the Matsubara frequencies while the filled circles denote the poles
of $F(R,\omega)$.}
\label{zmuc}
\end{center}
\end{figure}
Choosing the contour shown in Figure \ref{zmuc}, the expression above can be
evaluated in the complex $\omega$-plane as
\begin{eqnarray}
\nonumber
 \epsilon a^4&=&\frac{1}{\pi N^3 }\sum_{p_j}\left[2\pi i\sum_{i}
 \text{Res}~F(R,\omega_i)\right.\\
\nonumber
&-&\int_{\pi-i\theta}^{\pi}F(R,\omega)d\omega-
\int_{\pi}^{-\pi}F(R,\omega)d\omega\\
&-&\left.\int_{-\pi}^{-\pi-i\theta}F(R,\omega)d\omega-
\int_{-\pi}^{\pi}F(\omega)d\omega\right]~.~
\end{eqnarray}
The second and fourth terms cancel since $F$ satisfies
$F(R,\pi+i\eta)=F(R,-\pi+i\eta)$. Hence, we obtain
\begin{eqnarray}
\label{eqn:emuz}
\nonumber
\epsilon a^4&=&\frac{1}{\pi N^3}\sum_{p_j}\left[2\pi R_1\Theta\left(\frac{K(\hat{\mu})-L(\hat{\mu})}{2}-\sqrt{f}\right)\right.\\
&+&\left.\int_{-\pi}^{\pi}F(R,\omega)d\omega-\int_{-\pi}^{\pi}F(\omega)d\omega\right]~,~
\end{eqnarray}
where -i$R_1$ is the residue of the function $F(R,\omega)$ at the pole
$-i$~sinh$^{-1}(\sqrt{f}/R)$ .  It is clear from Eq. (\ref{eqn:emuz}) that
the condition $R = 1$ cancels the integrals, yielding the canonical Fermi
surface form of the energy density.  For $R\neq1$, there will in general be
violations of the Fermi surface on the lattice.  Moreover, in the continuum
limit $a \to 0$, one will in general have the $\mu^2$-divergences for
$R\neq1$ in the energy density.  The condition to obtain the correct
continuum values of $\epsilon=\mu^4/4\pi^2$ turns out to be
$K(\hat{\mu})-L(\hat{\mu})=2\hat{\mu} +O(\hat{\mu}^2)$.  That this
effective domain wall fermion satisfy the same condition as the overlap
\cite{bgs} suggest that such condition may be generically true for
Ginsparg-Wilson fermions. Also that one obtains identical condition in the
staggered case\cite{gav} suggests that the behavior near the continuum
limit dictates this condition.  Note also that the form used by Bloch and
Wettig \cite{wettig}, namely, exp($\pm \hat{\mu}$) for $K$, $L$, also
satisfies the condition $R = K \cdot L = 1$.

\begin{figure}
\begin{center}
 \includegraphics[scale=0.6]{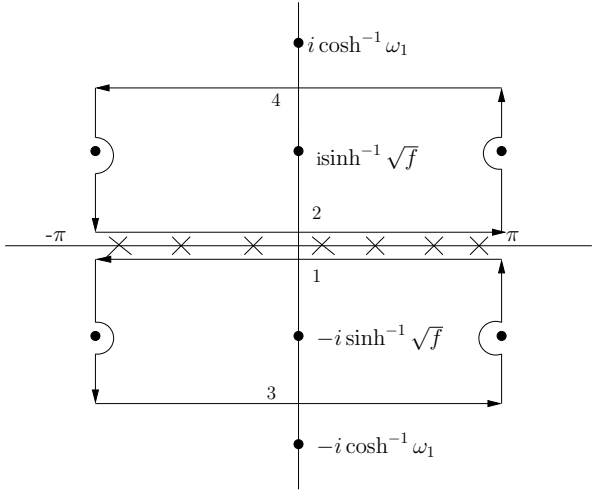}
\caption{Contour chosen for evaluating the energy density at finite 
temperature. The crosses indicate the Matsubara frequencies  
while the filled circles denote the poles of $F(\vec{p},\omega)$.}
\label{econt}
\end{center}
\end{figure}

\subparagraph{$T \ne 0$ :}
In order to choose the appropriate contour in the $T \ne 0$ case, note
that the function $F(R, \omega,\vec{p}) $ has poles at
$\cos^{-1}(\sqrt{d^2-g})=\pm i\sinh^{-1}\sqrt{f}$.  These turn out
to contain the physical poles in the continuum limit.  However, there are
additional(unphysical) poles at $\cos^{-1}(-\sqrt{d^2-g})=\pm \pi \pm i
\sinh^{-1}\sqrt{f}$, $\pm\pi\pm i\cosh^{-1}\frac{d}{2g}(\pm
i\cosh^{-1}\frac{d}{2g})$ for $\frac{d}{2g}>0(<0)$ and at $\pm
i\cosh^{-1}\omega_1$ where $\omega_1=\frac{d+4-4g}{2(g-2)}$.  The
definitions of the quantities $d$,$f$,$g$ are 
\begin{eqnarray}
\label{eqn:aux}
\nonumber
g&=&M-4+b, ~{\rm with}\\
\nonumber
b&=&\cos(a p_1)+\cos(a p_2)+\cos(a p_3)\\
\nonumber
f&=& h_1^2+h_2^2+h_3^2\\
\nonumber
d&=& 4+(M-4)^2+2(M-4) b +c, ~{\rm with} \\
c &=& \sum_{i<j< 4} 2 \cos(a p_i) \cos(a p_j)~.~
\end{eqnarray}
Note the obvious similarity with the  case of the overlap
fermions\cite{bgs}, where the same set of poles contribute as well.  Unlike
in the overlap case, however, there are no branch cuts present in this
case.  The contour chosen for evaluating the frequency sum shown in Figure
\ref{econt}, is thus different from that chosen for overlap fermions. The
residue of the pole enclosed by the contour for $F$ comes out to be,
\begin{equation}
 4\frac{\sqrt{f}}{\sqrt{1+f}}+\frac{\sqrt{1+f}-1}{\sqrt{f(1+f)}}G(M).
\end{equation}
with the first term yielding the continuum value of the energy density in
the limit of vanishing lattice spacing $a$. The energy density expression
comes after performing the contour integral comes out to be,
\begin{eqnarray}
\label{eqn:lat1}
\nonumber
\epsilon a^4& =&\frac{1}{N^3} \sum_{p_j}\left [4\frac{\sqrt{f}}{\sqrt{1+f}}
+\frac{\sqrt{1+f}-1}{\sqrt{f(1+f)}}G(M)
\right]\\
&\times&\frac{1}{e^{ N_T\sinh^{-1}\sqrt{f}}+1}
+\epsilon_3+\epsilon_4~,~
\end{eqnarray}
which again turns out to be similar to the overlap case.  Due to a
different functional form of $F$ and a different choice of contour, the
corresponding lattice correction terms $\epsilon_3,\epsilon_4$ which are
the line integrals of $F$ along lines 3,4 in the Figure \ref{econt}, are
different, leading to different finite size corrections.  In the continuum
limit the unphysical poles are pushed to infinity and the values of
$\epsilon_3,\epsilon_4$ vanish, leaving only the contribution of the
physical poles to the energy density: In the square bracket, only first
term gives the usual continuum expression with the other term vanishing
as $a\rightarrow0$.  The same treatment goes through in presence of
$\hat{\mu}$ only the contour has to be shifted along the imaginary $\omega$
plane by an amount dependent on $\hat{\mu}$ with the position of the poles
in the complex $\omega$-plane remaining unchanged. 
 
\subsection{$N_5\rightarrow\infty~,~a_5\rightarrow0~,~L_5=N_5 a_5=\text{finite}$}
In the case when the lattice spacing in the fifth direction $a_5\rightarrow
0$ and the number of sites $N_5\rightarrow \infty$ such that $L_5=N_5a_5$
is finite, the effective domain wall operator reduces to 
\begin{equation}
D_{DW}=(1+\hat{m_q})+(1-\hat{m_q})\gamma^5\text{tanh}(\frac{L_5}{2}
\gamma^5 D_{W}) 
\end{equation} 
Starting from the above expression we recover the overlap operator when
$L_5\rightarrow\infty$. With this effective domain wall operator, the
energy density can be evaluated\cite{gs} as,
\begin{eqnarray}
\label{eqn:effen}
\nonumber
&&\varepsilon a^4=\sum_{p_j,n}\frac{4 \sinh[\frac{sL_5}{2}]( \left(-h_4 h_5 \alpha
 +h^2 \gamma \right)  \cosh[\frac{sL_5}{2}]}{sN^3N_T}\\\nonumber
&&\frac{+\left(h_4 h_5 \alpha +(h_5^2+s^2)\gamma+2h_5 s^2 t\right)\cosh
(\frac{3 s L_5 }{2})- 2 s\sinh[\frac{s L_5}{2}] }{(h^2+\left(s^2+ h_5^2\right) \cosh[2 sL_5]}\\\nonumber 
 &&\frac{  (h_5^2 t+h_5\gamma +(h^2 t+h_4 \alpha +2 h_5(h_5 t+\gamma)) \cosh[ sL_5]) 
)}{-2 h_5 s \sinh[2 s L_5])}~,\\
\end{eqnarray}
where $\alpha$ and $\gamma$ are the same as defined previously and $t$ is
now defined as,
\begin{equation}
 t=\frac{(-\sin^2 ap_4+h_5 \gamma)(-\tanh \frac{ L_5 s}{2}+\frac{L_5 s}{2}
\text{sech} ^2 \frac{L_5 s}{2})}{s^2\tanh \frac{L_5 s}{2}}~.~
\end{equation} 
It was checked that the overlap energy density is obtained back 
when $L_5\rightarrow\infty$. We use the expression above for our numerical
work presented in section III.
 
\subsection{Finite $N_5$ and $a_5$}
While performing Monte Carlo simulations with domain wall fermions one
needs to work on lattices with finite number of sites in the fifth
dimension.  For finite $N_5$, the chiral symmetry is broken and it is
important to ascertain the dependence of the correction terms with $N_5$.
Evaluating the matrix ${\rm tanh} (N_5/2 \ln |T|)$ in Eq. (\ref{eqn:tanh}) 
various thermodynamic quantities of free domain wall fermions on the lattice
can be evaluated.  The energy density in the  massless limit  then is
\begin{eqnarray}
\label{eqn:ee}
\nonumber
&&\epsilon a^4=\frac{2}{N^3 N_T}\sum_{p_j,n}\left(\frac{ t^{'}}{t}+
\frac{2^{N_5}u^{'}}{2 ^{2 N_5+1}+2^{N_5} u}\right.\\
&&\left.~~~~~~~~~~~ -\frac{t u^{'}+u t^{'}-x q^{'}-(q-2)x^{'}} {tu-(q-2)x}\right)
\end{eqnarray}
where the quantities $u$ and $x$ are functions of h's defined in
 Eq. (\ref{eqn:hdef}) defined as,
\begin{eqnarray}
 \label{eqn:def}
\nonumber
&&u=\left(\frac{t-q}{h_5-1}\right)^{N_5}+\left(\frac{t+q}{1-h_5}\right)^{N_5}~,\\
&&x=\left(\frac{t-q}{h_5-1}\right)^{N_5}-\left(\frac{t+q}{1-h_5}\right)^{N_5}~.
\end{eqnarray}
The partial derivatives of the above variables are represented as the same
variables with a dash, and are functions of h's, $\alpha$ and
$\gamma$. 
\begin{eqnarray}
\label{eqn:deriv}
\nonumber
q^{'}&=&2 ss^{'}-2\gamma\\\nonumber
u^{'}&=&N_5\left(\frac{t-q}{h_5-1}\right)^{N_5-1}\left[
\frac{t^{'}-q^{'}}{h_5-1}-\frac{\gamma(t-q)}{(h_5-1)^2}\right]\\\nonumber
&+&N_5\left(\frac{t+q}{1-h_5}\right)^{N_5-1}\left[
\frac{t^{'}+q^{'}}{1-h_5}+\frac{\gamma(t+q)}{(1-h_5)^2}\right]~,\\\nonumber
x^{'}&=&N_5\left(\frac{t-q}{h_5-1}\right)^{N_5-1}\left[
\frac{t^{'}-q^{'}}{h_5-1}-\frac{\gamma(t-q)}{(h_5-1)^2}\right]\\\nonumber
&-&N_5\left(\frac{t+q}{1-h_5}\right)^{N_5-1}\left[
\frac{t^{'}+q^{'}}{1-h_5}+\frac{\gamma(t+q)}{(1-h_5)^2}\right]~.~
\end{eqnarray}
Again, we shall use these expressions for obtaining the numerical results
presented below where we also show the results for quark number susceptibility.
The same set of formulae remain valid for calculation of the susceptibility
except for the fact that $\alpha_{\mu}$ and $\gamma_{\mu}$ then are the
derivatives with respect to $\hat{\mu}$ and are defined as,
\begin{eqnarray}
\label{eqn:muderiv}
\nonumber
\alpha_{\mu}&=&\frac{\partial h_4}{\partial \hat{\mu}}=
\frac{ia}{a_4} \cos(a_4p_4-i\hat{\mu})~,~\\\nonumber
\gamma_{\mu}&=&\frac{\partial h_5}{\partial \hat{\mu}}=-ih_4\text{(for number density)}
\end{eqnarray}

\section{Numerical results for $\hat{\mu}=0$}

\subsection{$N_5=\infty,a_5=1$}
\begin{figure}
\begin{center}
 \includegraphics[scale=0.6]{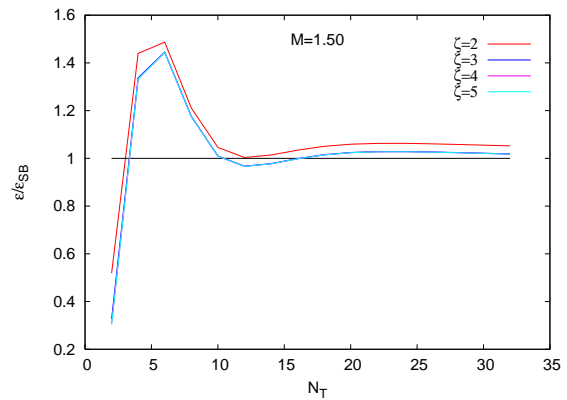}
\caption{The $\zeta$ dependence of the energy density of domain wall fermions
for $M=1.50$, $a_5=1$ and in the limit $N_5\rightarrow\infty$.}
\label{dwzeta}
\end{center}
\end{figure}

\begin{figure}
\begin{center}
 \includegraphics[scale=0.6]{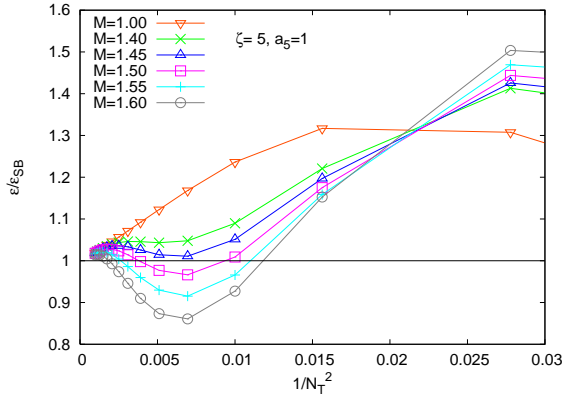}
 \caption{The variation of energy density of domain wall fermions with M in
 the limit $N_5\rightarrow\infty$ and $a_5=1$.}
\label{dwsgn}
\end{center}
\end{figure}

The goal of our numerical study is to find the optimum range of $M$ for
which the finite lattice spacing corrections are minimum and compare it
with that for the Dirac-Neuberger case \cite{bgs}.  We do this in the
chiral limit and set $\hat{m}_q = 0$.  The lattice energy density given by
Eq. (\ref{eqn:sgne}) was computed numerically by summing over the momenta
along the spatial and temporal directions.  The zero temperature part of
the energy density was determined in the limit $N_T\rightarrow\infty$ on a
lattice with a very large spatial extent $N$ by numerically evaluating the
$ap_4=\omega$ integral.  Holding the physical volume constant in units of
$T$ by keeping $V^{1/3}T = N/N_T \equiv \zeta$ fixed, we define the
continuum limit by $N_T \rightarrow \infty$.  The thermodynamic limit is
then achieved in the limit of large $\zeta$.   We first determine the
acceptable range of $\zeta$ by looking for $\zeta$-independence.  The
$\epsilon$ obtained by subtracting the zero temperature part from the
lattice energy density expression was normalized by its continuum value
$\epsilon_{SB}$. Figure \ref{dwzeta} displays the ratio
$\epsilon/\epsilon_{SB}$ as a function of
$N_T$ for different values of $\zeta$ at a fixed $M=1.50$. One notices
that for $\zeta\geq3$  the energy density plots lie on top of each other,
suggesting the thermodynamic limit to have reached  by $\zeta = 4-5$.
In order to highlight the deviations in the continuum limit, the same ratio
is exhibited for different M values for $\zeta=4$ in Figure \ref{dwsgn} as
a function of $1/N_T^2$ and listed in Table \ref{dwsgne} for a range of
$N_T$ likely to be used in simulations.  We choose to define the optimum
range of M as the values of M for which the thermodynamic quantities are
within 3\% of the continuum values for the smallest possible $N_T$.  One
sees from both the Figure \ref{dwsgn} and the Table \ref{dwsgne} that the
order $1/N_T^2$ corrections are minimum for M between 1.45-1.50 and
$N_T\geq12$.
\begin{table} 
\caption{$\epsilon/\epsilon_{SB}$ values for different M for $\zeta=4$}
\begin{tabular}{@{\extracolsep{\fill}}|c||c|c|c|c|c|}
\hline
$N_T$& M=1.0 & 1.40 & 1.45 & 1.50 & 1.55\\
\hline
4 & 0.909 & 1.240 & 1.286 & 1.333 & 1.384\\
6 & 1.308 & 1.413 & 1.426 & 1.444 & 1.469\\
8 & 1.317 & 1.221 & 1.197 & 1.175 & 1.159\\
10 & 1.236 & 1.090 & 1.051 & 1.009 & 0.966\\
12 & 1.168 & 1.048 & 1.011 & 0.966 & 0.915\\
14 & 1.123 & 1.043 & 1.014 & 0.977 & 0.930\\
16 & 1.092 & 1.046 & 1.026 & 0.998 & 0.960\\
\hline
\end{tabular}
\label{dwsgne}
\end{table}
The correction terms for $M=1$ are  linear in $1/N_T^2$ for $N_T \ge 10$
and are about $20\%$ of the continuum value even for $N_T=12$.  This is
similar to that reported earlier for the overlap fermions\cite{bgs}.
Though the continuum extrapolation with $M=1$ is easier due to the linear
functional form, it is computationally expensive, needing simulations at
more values of $N_T$, each greater than 10.  Thus $M=$1.45-1.50 seems to be
an optimum range for lattice simulation of the energy density of domain
wall fermions.  We have also varied the lattice spacing along the fifth
dimension $a_5$ to find out how the cut-off dependent terms change with
it. The correction terms to the energy density for $a_5=0.5$  at small
lattice sizes $N_T\leq10$ are indeed larger than that for $a_5=1$ for the
above mentioned optimum range but for $N_T>12$ such terms are again within
2-3\% of the Stefan Boltzmann value.  The optimum $M$ range for which the
lattice artifacts are minimum shifts to 1.50-1.60. Thus there is a marginal
dependence on $a_5$ for $N_T\geq10$.  Reducing $a_5$ further does not
increase the range much as we demonstrate in the plot for $a_5\rightarrow0$
in Figure \ref{dwtanh}.

\subsection{$N_5\rightarrow\infty~,~a_5\rightarrow0~,~L_5=N_5 a_5=\text{finite}$}

\begin{figure}
\begin{center}
 \includegraphics[scale=0.6]{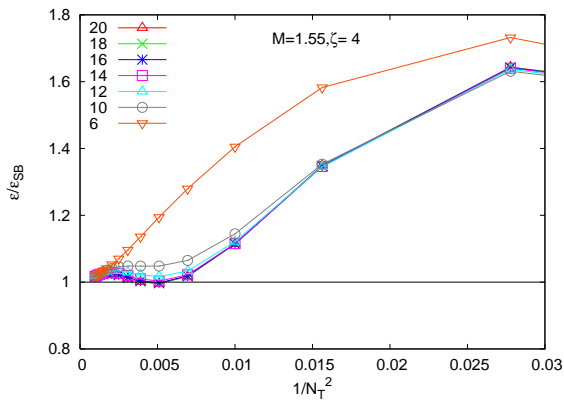}
\caption{The variation of energy density on lattice with $N_T$ for
 domain wall fermions for different $L_5$, as shown by the respective
labels, and $M=1.55$.}
\label{L5dep}
\end{center}
\end{figure}

\begin{figure}
\begin{center}
\includegraphics[scale=0.6]{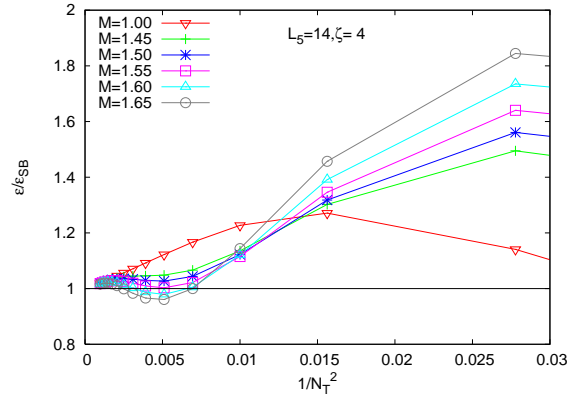}
\caption{The variation of energy density on lattice with $1/N_T^2$ for
 domain wall fermions for $L_5=14$ and different M.}
\label{dwtanh}
\end{center}
\end{figure}

Next we  investigated the limit $N_5\rightarrow\infty~,~a_5\rightarrow0$
such that $L_5=N_5 a_5=\text{finite}$ in order to estimate numerically the
value of $L_5$ for which we recover the overlap energy density starting
from Eq. (\ref{eqn:effen}). As can be observed from Figure \ref{L5dep},
$L_5$-independent results are obtained for $L_5 \ge 14$ for $M=1.55$.  This
was also the case for a range of $M$ around this value.  For $L_5\leq10$
the convergence towards the $\epsilon_{SB}$ value was seen to be very slow
for all M and we find that the continuum value is not reached even for
lattice size as large as $N_T=32$.  Figure \ref{dwtanh} displays the
results as a function of $1/N_T^2$ for $L_5 =14$ and various values of $M$
indicated on it.  The deviations from the continuum for such $L_5$ are less
than 3 \% for the range of M between 1.50-1.60, in agreement with the
overlap results \cite{bgs}.

\subsection{Finite $N_5$ and $a_5=1$}
\begin{figure}
\begin{center}
\includegraphics[scale=0.6]{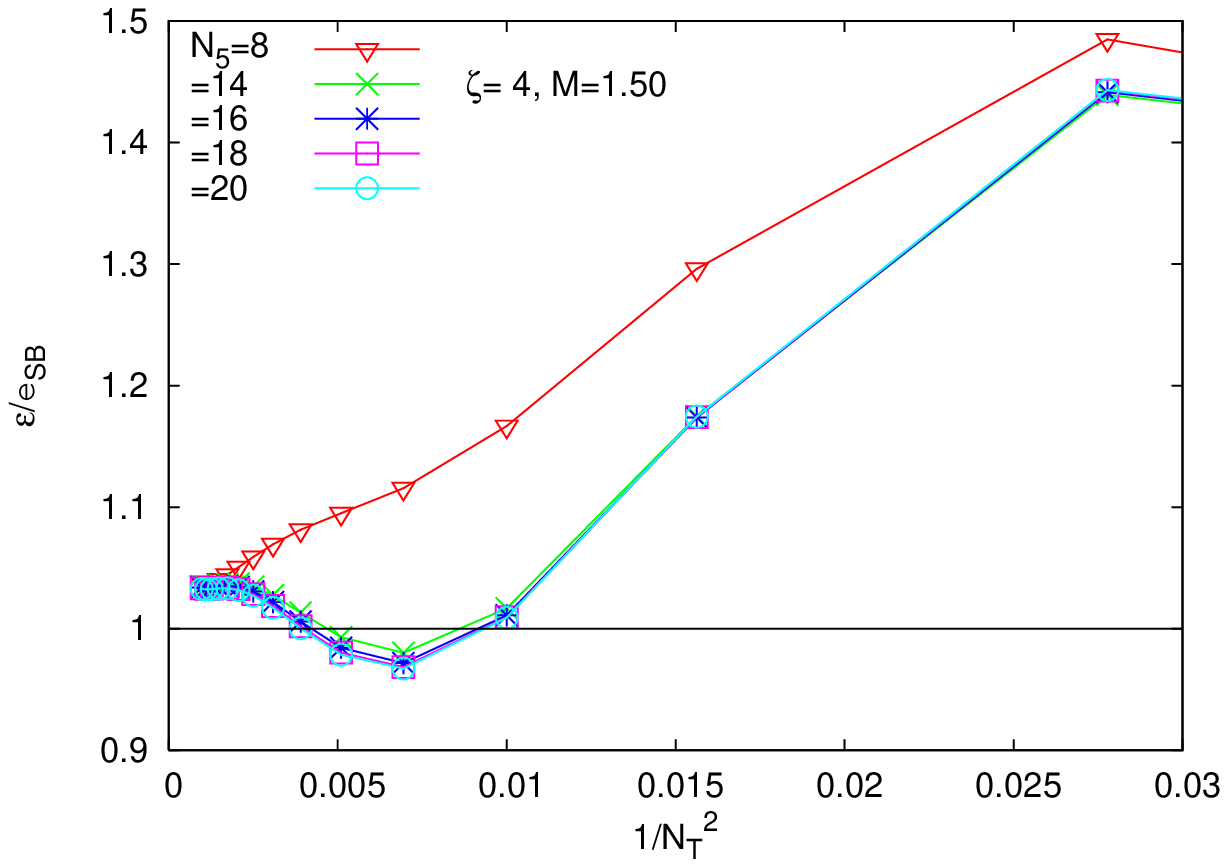}
\includegraphics[scale=0.6]{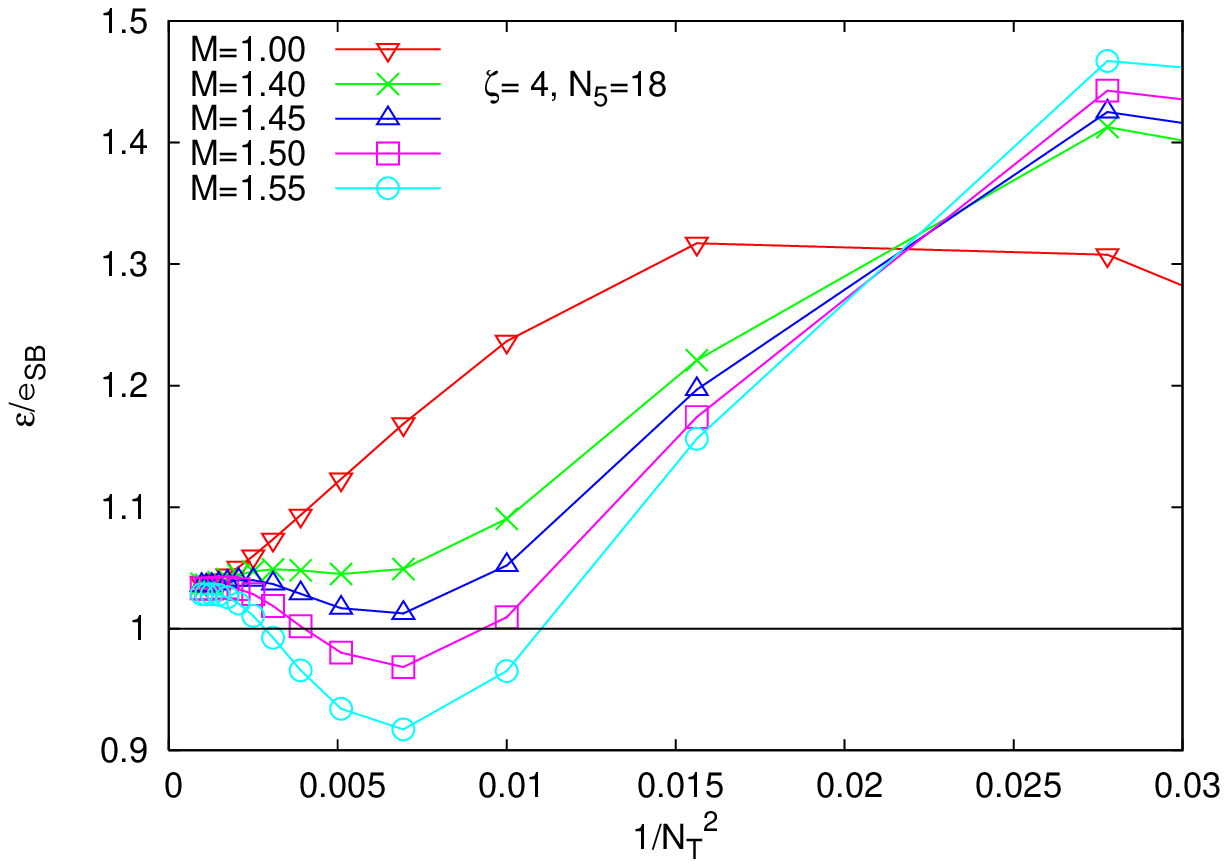}
\caption{The variation of energy density on lattice with $1/N_T^2$ for domain
 wall fermions at different $N_5$ for $M=1.50$ (upper panel) and at $N_5=18$
 for different $M$(lower panel).}
\label{dwfin}
\end{center}
\end{figure}

The case of finite $N_5$ with $a_5 =1$ is clearly of most interest for
practical simulations with dynamical fermions.  Earlier numerical studies
for free domain wall fermions\cite{flem,hegde} employed $M=1.0$ and found
somewhat slow convergence of various thermodynamic quantities towards their
continuum values. We intend to check whether tuning the value of M results
in a faster convergence. For that purpose we have computed the energy
density expression for finite $N_5$ and $a_5=1$ in Eq.(\ref{eqn:ee}) by
summing over all the discrete momenta.  We display those results for
$\epsilon/\epsilon_{SB}$ in Figure \ref{dwfin}.  The upper panel shows the
results for a series of $N_5$ and a fixed $M=1.5$.  The results are seen to
become $N_5$-independent by $N_5=18$, making it an optimum choice for
obtaining continuum results on the lattice.  The lower panel shows the
$M$-variation for $N_5 =18$. Table \ref{dwen} provides the values for
lattices with reasonable $N_T$-extent.   The general trend is clearly the
same as above with $M=$1.45-1.50 emerging as the range for which the
Stefan-Boltzmann limit is reached to within 3-4\% for $N_T\geq10$ (Table
\ref{dwen}).  Interestingly, $N_5 =18$ seems to mimic the $N_5 \rightarrow
\infty$ limit quantitatively rather well as can be seen by comparing the
Tables \ref{dwsgne} and \ref{dwen}.   Consequently, the same optimum range
of $M$ is obtained for both.
\begin{table}
\caption{$\epsilon/\epsilon_{SB}$ values for different M for $\zeta=4,N_5=18$}
\begin{tabular}{@{\extracolsep{\fill}}|c||c|c|c|c|c|}
\hline
$N_T$& M=1.0 & 1.40 & 1.45 & 1.50 & 1.55\\
\hline
4 & 0.909 & 1.240 & 1.285 & 1.333 & 1.383\\
6 & 1.308 & 1.413 & 1.425 & 1.443 & 1.467\\
8 & 1.317 & 1.221 & 1.197 & 1.174 & 1.156\\
10 & 1.237 & 1.090 & 1.052 & 1.009 & 0.965\\
12 & 1.169 & 1.049 & 1.013 & 0.968 & 0.917\\
14 & 1.123 & 1.045 & 1.017 & 0.980 & 0.934\\
16 & 1.093 & 1.048 & 1.029 & 1.002 & 0.966\\
\hline
\end{tabular}
\label{dwen}
\end{table}

\section{Numerical results for $\hat{\mu}\neq0$}

It should be noted that in this case $T$ is no longer Hermitian but as long
as the condition given in the Eq. (\ref{eqn:cond}) is satisfied the
effective operator in Eq. (\ref{eqn:sgne}) is well defined.  We shall
restrict the range of $\hat{\mu}$ to ensure that it is so.  We choose $K$ and
$L$ to be $e^{\pm \hat{\mu}}$ respectively in our numerical computations as
suggested in \cite{wettig}.   Our aim again is to  find the optimum $M$ for
which the continuum results are obtained with least computational effort,
and compare it with our the range obtained from the energy density above.
We consider two observables here. One is the change in the energy density
due to nonzero $\mu$ : $\Delta
\epsilon(\mu,T)=\epsilon(\mu,T)-\epsilon(0,T)$. In the continuum limit this
is 
\begin{equation}
 \frac{\Delta \epsilon(\mu,T)}{T^4}=\frac{\mu^4}{4 \pi^2
T^4}+\frac{\mu^2}{2 T^2}.
\label{eqn:DelEmu}
\end{equation}
Another observable we studied was the quark number susceptibility at
$\hat{\mu}=0$. It is defined for any $\hat{\mu}$ by,
\begin{equation}
\chi=\frac{1}{N^3 a^2 N_T}\left(\frac{\partial^2 \ln \det D}{\partial
\hat{\mu}^2  }\right)_{a_4}~,~
\end{equation}
and in the continuum is given by,
\begin{equation}
\label{eqn:sus}
\chi(\mu)=\frac{\mu^2}{\pi^2}+\frac{T^2}{3}~.~ 
\end{equation}
We will focus on $\chi(0)$ here due to its importance in the applications
to the heavy ion collisions.

\begin{figure}
\begin{center}
 \includegraphics[scale=0.6]{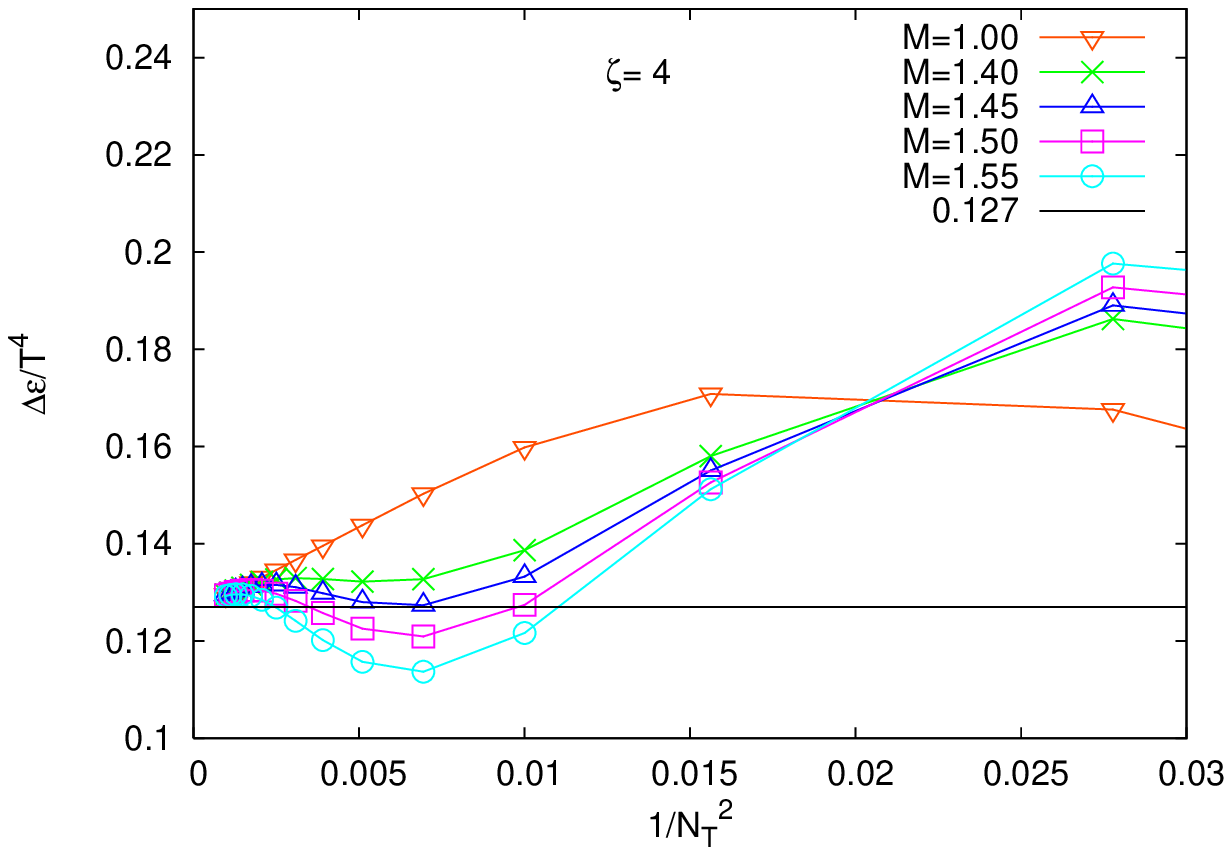}
 \includegraphics[scale=0.6]{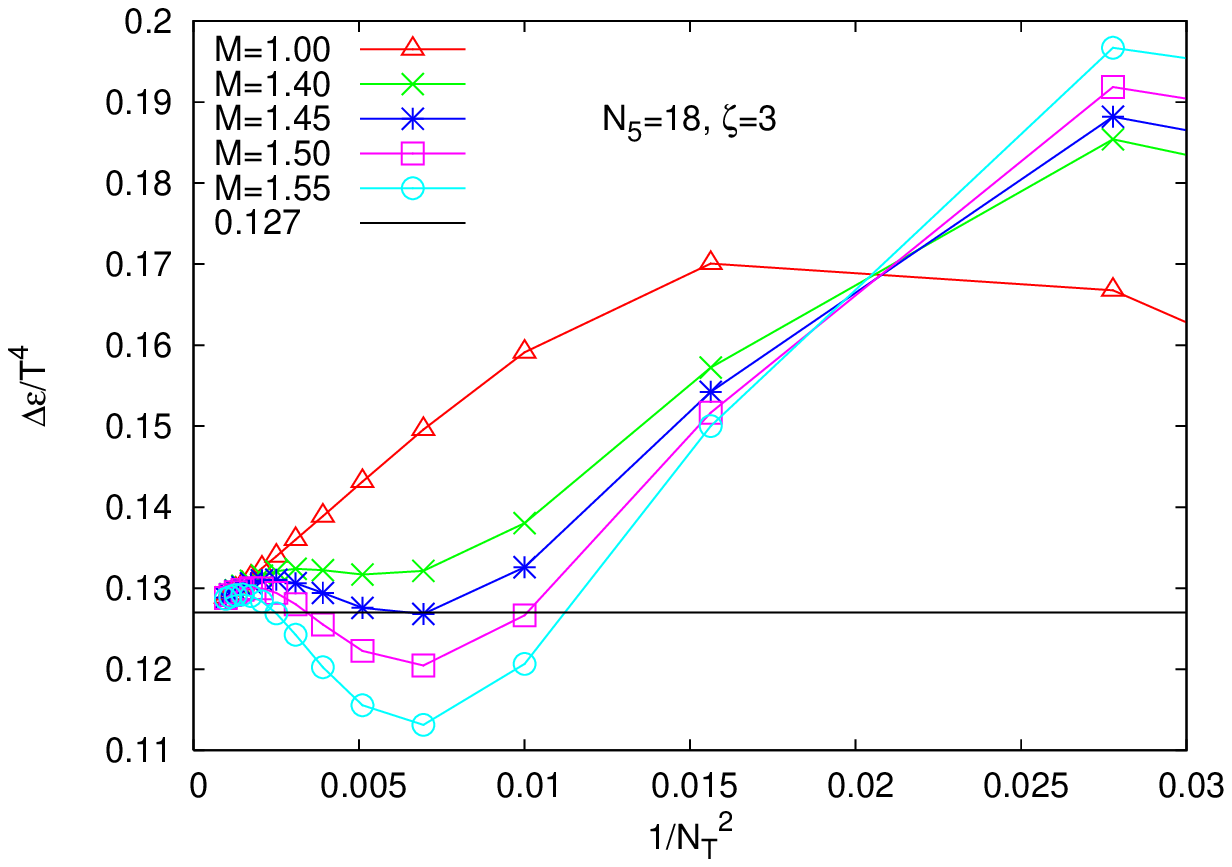}
\caption{The energy density of domain wall fermions in presence of $\hat{\mu}$ 
for different $M$ for $\hat{\mu}=0.5/N_T$ and infinite $N_5$(upper panel) 
and for $N_5=18$ (lower panel).}
\label{dwsgndelE}
\end{center}
\end{figure}

We estimated numerically $\Delta \epsilon(\mu,T)$ for $\mu/T=\hat{\mu}N_T$
fixed at 0.5.  The upper and lower panels of the Figure \ref{dwsgndelE}
display our results for this observable in the units of $T^4$ for $N_5 =
\infty$ and 18 respectively for the $M$ values indicated.  The horizontal
line in each case shows the expected result in the continuum limit from
Eq.(\ref{eqn:DelEmu}).  From the Figure \ref{dwsgndelE} it is evident that
there are no $\mu^2$ divergences on the lattice, as expected.  The
deviations from the continuum limit are due to the $M$ dependent finite
size effects. These correction terms are again seen to be small for the
same optimum range of $1.45\leq M\leq1.50$ for both the cases, as obtained
in the zero chemical potential case in section III.

\begin{figure}
\begin{center}
 \includegraphics[scale=0.6]{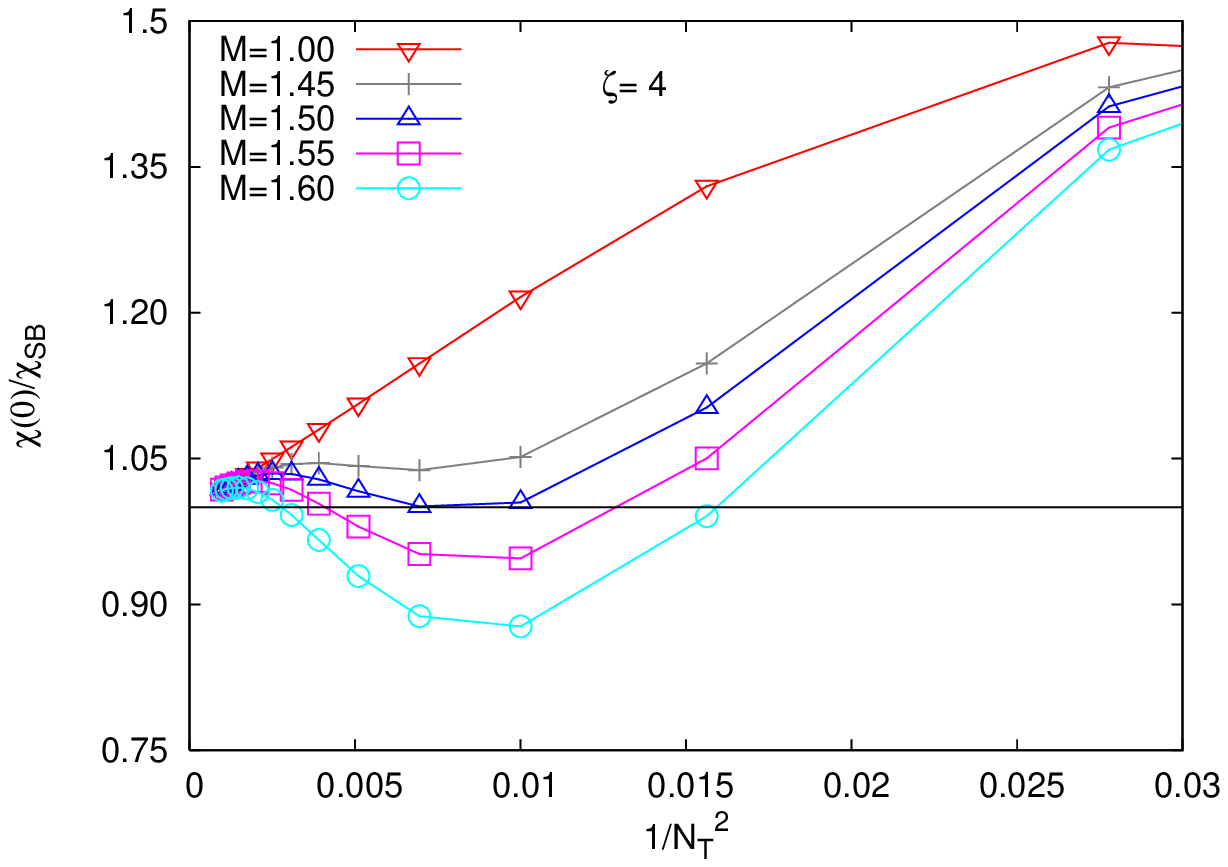}
\includegraphics[scale=0.6]{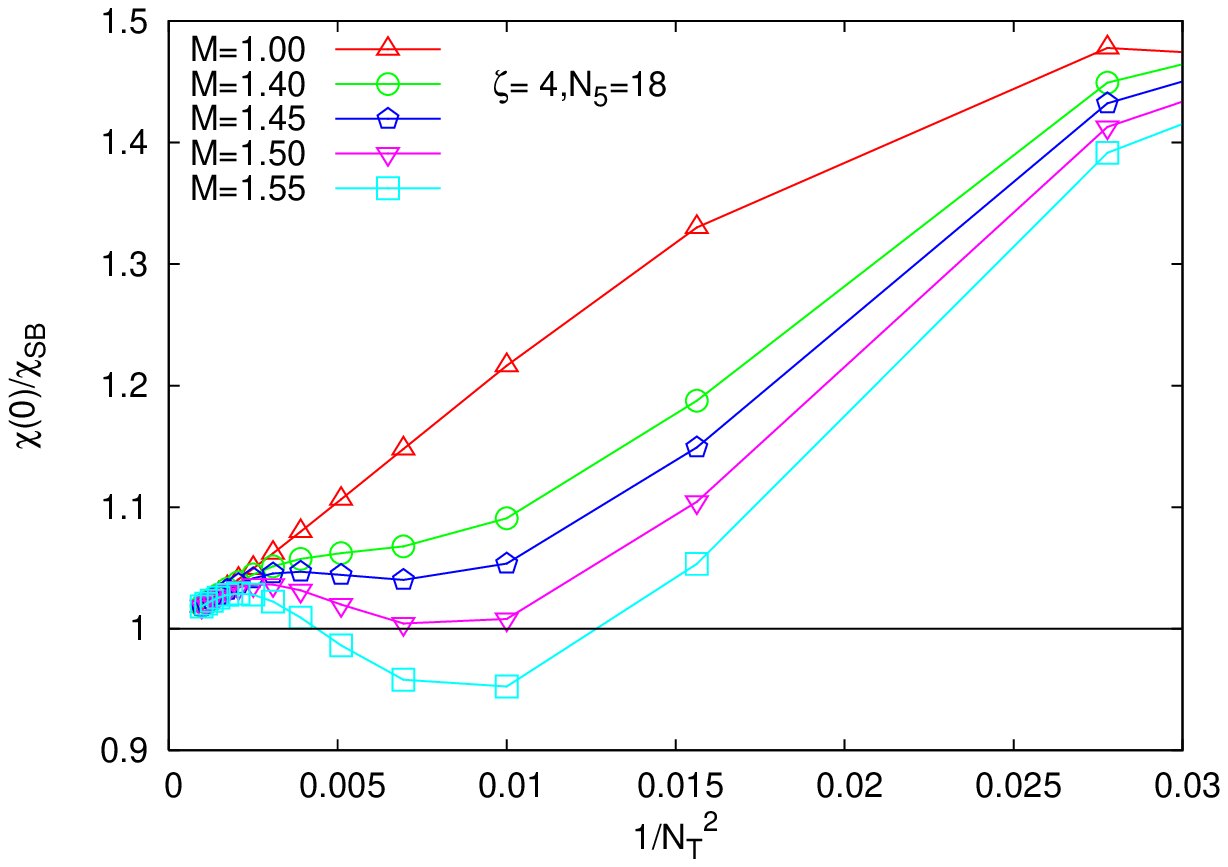}
\caption{ The quark number susceptibility as a function of $1/N_T^2$ for
$M$ values as indicated for $\zeta =4$ and $N_5 = \infty$ (upper panel)
and 18 (lower panel).}
\label{dwchi}
\end{center}
\end{figure}

The $N_5$ dependence of the quark number susceptibility at $\hat{\mu}=0$ 
is plotted in Figure \ref{dwchi}. It too exhibits a convergence to the 
infinite $N_5$ results for $N_5\geq16$, indicating that $N_5 = 18$ can
again be used safely to approximate the infinite $N_5$.  Figure \ref{dwchi}
shows the $M$-dependence of the quark number susceptibility at
$\hat{\mu}=0$.  Both the $N_5 = \infty$ (upper panel) and 18 (lower panel)
show small deviations from the Stefan-Boltzmann value of $1/3$ for
 $1.45 \le M \le 1.55$ range and for $N_T\geq10$. 
\begin{figure}
 \includegraphics[scale=0.6]{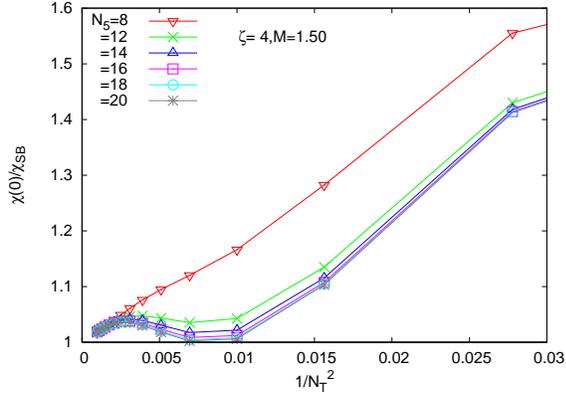}
\begin{center}
\caption{The quark number susceptibility as a function of $1/N_T^2$ for
$N_5$ values as indicated  for $M=1.5$ and $\zeta =4$.}
\label{dwchiN5 }
\end{center}
\end{figure}
Recent computations of this susceptibility\cite{fks} for the interacting
domain wall fermions were performed with $M=1.8$. Of course, one expects
some shift in $M$ due to the interactions, which should however be small
for large enough temperature where one expects those computations to
approach the free quark gas results. In all our plots we find that for the
optimum $M$ range, the deviations from the ideal gas results at smaller
$N_T=$4-8 are quite significant but with a relative mild $M$-dependence for
$M> 1.4$.  Thus a slightly larger value of $M$ than the optimum range we
found may not change the finite size effects drastically for small $N_T$.
What one does need to be careful about though is the extrapolation to the
continuum limit.  For the optimal range of $M$ and $N_T \ge 10$, the
smallness of corrections compared to other errors in the computations may
make it a less important issue.

\section{Improvement of the chiral fermion kernels}

In the previous sections we observed that the fermions with exact chiral
symmetry on the lattice have large $1/N_T^2$ corrections for small $N_T$.
While we found that the continuum limit for various thermodynamic quantities
can be approached faster by choosing the irrelevant parameter $M$ in the range
1.45-1.55,  the correction terms for $N_T=$4-6 are about twice that of the
Stefan-Boltzmann result for such a choice of $M$ too.  Here we describe our
attempts to improve the convergence to the continuum results for small $N_T$
and even for $M=1.0 $.  Having the option of the choice of $M=1.0$ may be
useful since it has been noted previously\cite{vranas,cap} that the residual
mass for such a choice of $M$ is zero for a range of $N_5$ at the tree level.

\subsection{Domain wall kernel}
The domain wall operator given in Eq. (\ref{eqn:tanh}) is a matrix-function
of the Wilson-Dirac operator as in Eq. (\ref{eqn:wilop}). It is clear that
its improvement may lead to a better domain wall operator, or indeed even
a better overlap operator, one is looking for. Inspired by the attempts to
improve the staggered fermions in the so-called Naik-action \cite{naik}, we
add three-link terms to the $D_W$ as below.
 \begin{eqnarray}
 \label{eqn:impwilop}
\nonumber
&&D_W(x,x',\hat{\mu})=\left(3+\frac{a}{a_4}-M\right)\delta_{x,x'}-\\\nonumber
&&\sum_{j=1}^{3}\left(U_j^{\dagger}(x-\hat{j})\frac{1+c_1\gamma_j}{2 }
\delta_{x,x'+\hat{j}}+U_j(x)\frac{1-c_1\gamma_j}{2 }\delta_{x,x'-\hat{j}}\right)\\\nonumber
&-&\frac{a}{a_4}\left(U_4^{\dagger}(x-\hat{4})\frac{1+c_1\gamma_4}{2 }
\delta_{x,x'+\hat{4}}+U_4(x)\frac{1-c_1\gamma_4}{2 }\delta_{x,x'-\hat{4}}\right)\\\nonumber
&-&\sum_{j=1}^{3}\left(U_j^{\dagger}(x-3\hat{j})\frac{c_3\gamma_j}{6 }
\delta_{x,x'+3\hat{j}}-U_j(x)\frac{c_3\gamma_j}{6}\delta_{x,x'-3\hat{j}}\right)\\\nonumber
&-&\frac{a}{a_4}\left(U_4^{\dagger}(x-3_{\hat{4}})\frac{c_3\gamma_4}{6}
\delta_{x,x'+3_{\hat{4}}}
-U_4(x)\frac{c_3\gamma_4}{6}\delta_{x,x'-3_{\hat{4}}}\right)\\
\end{eqnarray}
Comparing with the Eq. (\ref{eqn:wilop}), it is clear that the modification
amounts to replacing $\gamma_\mu$ by $(c_1 +c_3/3) \gamma_\mu$.  The
Wilson mass term, added to remove the doublers, is kept unchanged. 
Note that the modified $D_W$-operator is still $\gamma_5$-hermitian for
arbitrary real values of the coefficients $c_1$ and $c_3$. The new domain
wall operator can therefore be derived in the same way as Eq. (\ref{eqn:tanh})
 was obtained.  We fix the coefficients by demanding the
dispersion relation for free fermions on the lattice to be the same as in
the the continuum up to $O(a^4p_j^4)$.  We find that all the terms at 
$O(a^3p_j^3)$ are eliminated for the coefficients $c_1=9/8,~c_3=-1/8$. 
We employ them below for the calculation of the thermodynamic quantities.

\begin{figure}
\begin{center}
\includegraphics[scale=0.6]{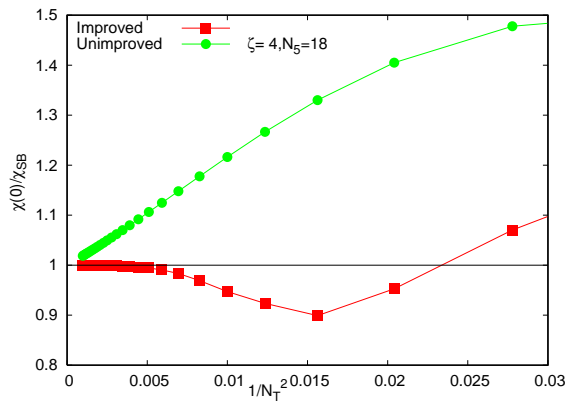}
\caption{The susceptibility of improved and the conventional domain wall
fermions at $M=1.0$ as a function of $1/N_T^2$. }
\label{dwimp}
\end{center}
\end{figure}

The ratio of quark number susceptibilities, $\chi/\chi_{SB}$, computed
using the above modified domain wall operator \cite{f1}, is plotted as a
function of $1/N_T^2$ as in Figure \ref{dwimp} along with that for the
unimproved domain wall operator of Eq. (\ref{eqn:tanh}). We used $M=1$,
$\zeta=4$, $N_5 =18$ and $a_5=1$ for this computation. One clearly notices
that the large correction terms ($\sim 45$\%) at $N_T=$6-8 for the usual
domain wall operator go down to about 7-8 \%. Indeed, the size of
corrections go down further as $N_T$ increases. Similarly, the the energy
density of such improved fermions also exhibited smaller, about 15-5\%,
deviations from the continuum for $N_T=$6-10, as compared to about 30 \%
in the lower panel of Figure \ref{dwfin}.

\subsection{Overlap kernel}

\begin{figure}
\begin{center}
 \includegraphics[scale=0.6]{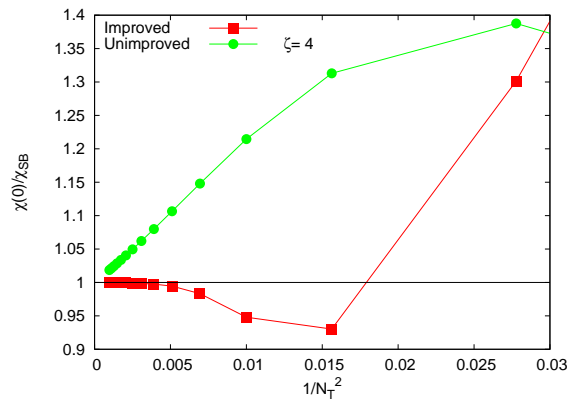}
\caption{The $\chi/\chi_{SB}$ as a function of $1/N_T^2$ for the improved
and the canonical overlap operators at $M=1.0$.}
\label{impoc}
\end{center}
\end{figure}

From section II, we know that the overlap operator can be derived as a
special limiting case of the domain wall operator. It would be thus
interesting to check how the improvement in the Wilson Dirac operator in Eq.
(\ref{eqn:impwilop}) fairs in the overlap case.  For that purpose we compute
the quark number susceptibility for non-interacting fermions on a $N^3\times
N_T$ lattice numerically with the corresponding improved overlap operator.
The $\chi/\chi_{SB}$ does have lower $1/N_T^2$ corrections for $N_T=6,8$
than for the conventional overlap  operator with $M=1$ as shown in Figure
\ref{impoc}. We also observe a faster approach to the continuum result with
such improved overlap operator than with the Neuberger overlap operator even
with optimum $M=1.55$ reported in \cite{bgs}. Another advantage is that the
thermodynamic quantities calculated from this improved operator are free
from oscillations at odd-even values of $N_T$ exhibited \cite{bgs} by the
usual overlap operator.  The improvement in the energy density is marginal
up to $N_T =8$ but substantial for $N_T =10$ onwards.

\section{Conclusions \& Discussions} 
 
Since the chiral violations vanish exponentially with the number of sites
$N_5$ in the fifth dimension, the domain wall fermions offer a more
practical alternative to the overlap fermions and yet have exact flavor and
spin symmetry.  We have computed the energy density and susceptibility at
zero chemical potential of such fermions numerically for both finite and
infinite $N_5$.  The chiral symmetry is exact in the latter case and a
choice $M$ between 1.45-1.50 allows faster convergence to the continuum
results. We have also verified analytically that the energy density has the
correct continuum value in the chiral limit.  Varying the number of lattice
sites in the fifth dimension, we have shown that $N_5=18$ is sufficient to
restore chiral symmetry. 

We found that introducing chemical potential $\hat{\mu}$ in domain wall
operator leads to chiral symmetry breaking even for infinite $N_5$. But if
we do allow that, there exist a large class of functions $K(\hat{\mu})$ and
$L(\hat{\mu})$, with $K(\hat{\mu}) \cdot L(\hat{\mu}) = 1 $, for which
there are no $\hat{\mu}$-dependent divergent terms in the physical
observables.  From the numerical evaluation of the energy density in
presence of $\hat{\mu}$ , we conclude that the optimum range of $M$ remains
the same. The lattice cut-off effects are however very large for small
$N_T$ = 4-8. By systematically removing the dominant correction terms
to the continuum value of the chiral fermion operators we have achieved a
faster convergence to the continuum as well as small $1/N_T^2$ corrections
for small lattice sizes even for $M=1.0$. This set of optimum parameters is
anticipated to produce similar results in full QCD simulations with chiral
fermions though an explicit check needs to be done.

\section*{Acknowledgments}
S.S would like to acknowledge the Council of Scientific and Industrial
Research(CSIR) for financial support.\\

\appendix

\section{Proof for GW relation}
\label{app1}
We show here that the effective domain wall operator Eq. (\ref{eqn:effdw})
satisfies the Ginsparg Wilson relation :
\begin{eqnarray}
\nonumber
 \gamma_5 D_{DW}&+&D_{DW}\gamma_5=\gamma_5(1-\gamma_5\epsilon(\ln|T|))\\\nonumber
&+&(1-\gamma_5\epsilon(\ln |T|))\gamma_5\\\nonumber
&=&2~\gamma_5-\epsilon(\ln |T|)-\gamma_5\epsilon
(\ln |T|)\gamma_5\\
\nonumber
D_{DW}\gamma_5 D_{DW}&=&(1-\gamma_5\epsilon(\ln |T|))\gamma_5(1-\gamma_5
\epsilon(\ln |T|))\\\nonumber
&=&\gamma_5-\gamma_5\epsilon(\ln|T|)\gamma_5-\epsilon(\ln |T|)+\gamma_5
\epsilon^2(\ln |T|)\\\nonumber
\text{Since,}\\\nonumber
 \epsilon(\ln |T|)&=&\frac{\ln |T|}{\sqrt{(\ln |T|) \ln |T|}}\\
\therefore\epsilon^2(\ln |T|)&=&\frac{\ln |T|\ln |T|}{(\sqrt{\ln |T|\ln |T|})^2}=1
\end{eqnarray}
Hence  $ \{ \gamma_5, D_{DW} \} = D_{DW}\gamma_5 D_{DW}$


\begin{thebibliography}{10}
\bibitem{f1}
Following \cite {gavlat}, we use $K_3(\hat{\mu}) = K^3(\hat{\mu})$
and $L_3(\hat{\mu}) = L^3(\hat{\mu})$ for introducing $\hat{\mu}$ for the 3-link
terms.
\end{thebibliography}
\end{document}